\documentclass[final]{elsarticle}

\usepackage{amssymb}

\journal{Medical Engineering \& Physics}
\usepackage{graphicx}
\usepackage{array}
\usepackage{mdwmath}
\usepackage{mdwtab}
\usepackage{eqparbox}
\usepackage{lineno}
\usepackage{url}
\usepackage{color}
\usepackage[font=footnotesize]{subfig}
\begin{document}

\begin{frontmatter}



\title{Screening~of~Obstructive~Sleep~Apnea~with\\
Empirical~Mode~Decomposition~of~Pulse~Oximetry}


\author[FIUNER,CONICET]{Gast\'on Schlotthauer\corref{gaston}}
\author[SINC,CONICET]{Leandro E. Di Persia\corref{leandro}}
\author[RESP]{Luis D. Larrateguy\corref{luis}}
\author[SINC,CONICET]{Diego H. Milone\corref{diego}}
\address[FIUNER]{Lab. of Signal Processing and Nonlinear Dynamics, Facultad de
Ingenier\'{\i}a, Universidad Nacional de Entre R\'ios, Argentina}
\address[CONICET]{National Council of Scientific and Technical Research
(CONICET) Argentina}
\address[SINC]{Research Center for Signals, Systems and Computational
Intelligence (sinc(i)), Facultad de Ingenier\'{\i}a y Ciencias H\'{\i}dricas,
Universidad Nacional del Litoral, Argentina}
\address[RESP]{Centro de Medicina Respiratoria, Paran\'a,~Argentina}
\cortext[gaston]{gschlotthauer@conicet.gov.ar}

\begin{abstract}

Detection of desaturations on the pulse oximetry signal is of great importance
for the diagnosis of sleep apneas. Using the counting of desaturations, an index
can be built to help in the diagnosis of severe cases of obstructive sleep
apnea-hypopnea syndrome. It is important to have automatic detection methods
that allows the screening for this syndrome, reducing the need of the expensive
polysomnography based studies. In this paper a novel recognition method based on
the empirical mode decomposition of the pulse oximetry signal is proposed. The
desaturations produce a very specific wave pattern that is extracted in the
modes of the decomposition. Using this information, a detector based on properly
selected thresholds and a set of simple rules is built. The oxygen desaturation index constructed from these detections produces a
detector for obstructive sleep apnea-hypopnea syndrome with high sensitivity
($0.838$) and specificity ($0.855$) and yields better results than standard
desaturation detection approaches.
\end{abstract}


\end{frontmatter}

\section{Introduction}
Sleep disorders include more than $80$ frequent pathologies in adults and
children \cite{AASM2001}. Such disorders cause daytime sleepiness, affecting
between $35$ and $40\%$ of the adult population of USA, and are an important
cause of morbidity and mortality. As a result of this high prevalence, severe
complications, and concomitant diseases in the non treated cases, there are very
important associated costs \cite{Hos2002}. The more common and important sleep
pathology is the obstructive sleep apnea-hypopnea syndrome (OSAHS). This
disorder is characterized by repetitive airflow reduction caused by an
intermittent partial or complete upper airway obstruction during sleep. The main
consequences of this disorder are sleep fragmentation, reduced blood oxygen
saturation, and excessive daytime somnolence \cite{Hor2007, Sal2007, Una2002,
Str1996}. According to recent studies
\cite{w_obstructive_2013,lurie2011obstructive}, the prevalence of OSAHS in a
general population, without taking into account symptoms of sleepiness, has been
estimated to be $24\%$ in a males and, when associated with these symptoms, it
decreases to  approximately $3-7\%$ in men and $2-5\%$ in women. It is worth to
be mentioned that it is much higher, e.g. $\geq 50\%$, in patients with cardiac
or metabolic disorders than in the general population.

The current gold standard for the diagnosis of OSAHS is polysomnography (PSG).
PSG is an overnight study made at a sleep center, in a quiet and dark room, that
consists of simultaneous recording of electroencephalography (EEG),
electrooculography (EOG), electromyography (EMG), electrocardiography (ECG),
oxygen saturation (SpO$_2$), oronasal airflow, thoracic and abdominal movement,
body position, and other signals. PSG allows to estimate the apnea/hypopnea
index (AHI) that is used as the primary index of OSAHS severity. PSG is
supervised by a technician, and its analysis requires a tedious scoring, often
by hand \cite{Thu2007}. This study is cost intensive, its availability is
limited, and only one study can be made per night.

As alternatives to PSG, several approaches have been proposed using cardiac,
respiratory, and snore sounds \cite{Yad2006,soler2012}, pulse oximetry
\cite{Hor2007}, ECG \cite{Roc2004}, nasal airway pressure
\cite{Sal2007,caseiro2010} and combinations of several signals \cite{Ray2003}.
These signals were studied by time-frequency analysis techniques \cite{Kan2005},
statistical approaches based on several \textit{ad hoc} indexes \cite{Ray2003},
empirical mode decomposition \cite{Sal2007,caseiro2010}, information theory
\cite{Hor2007}, linear and quadratic discriminants \cite{Yad2006}, and other
methods. Unlike other signals for which the recording instrumentation is more complex, nocturnal pulse oximetry is a low-cost technique and it can be easily applied in
outpatient studies with the purpose of screening of {OSAHS}. However, pulse oximetry requires more sophisticated processing tools to extract relevant information.

Empirical Mode Decomposition (EMD) is a complete data-driven signal analysis
technique, that can be applied to nonstationary and nonlinear signals, proposed
by Huang et al. \cite{Huang:1998} \footnote{Details about EMD can be found in the
Supplementary Material accompanying this paper.}. EMD decomposes a signal into a
usually small number of components known as Intrinsic Mode Functions (IMF) or
modes. EMD was successfully used for the extraction of the respiratory signal
from ECG \cite{Bal2004}, and for detecting apneas processing the nasal airflow
signal \cite{Sal2007} and even the ECG \cite{mendez2010automatic}. As an
undesired effect, we can mention the problem known as ``mode mixing'', where
very similar oscillations are present in different modes. This is partially
alleviated with noise-assisted EMD versions, as the Ensemble EMD \cite{wu2009}
with very good results in voice processing \cite{sch2009eusipco}, but with high
computational cost and a residual noise in the reconstructed signal. More recent
noise assisted versions overcome some of these problems
\cite{torres2011,colominas2012}.

In this work, we present an algorithm based on EMD for detecting desaturations
associated with sleep apnea/hypopnea in pulse oximetry signals. The purpose of
this procedure is to estimate an index that behaves in a similar way than the
classical apnea/hypopnea index derived from PSG, but using only information from
oxygen desaturations measured by pulse oximetry. This will be done by
decomposing the oximetry signal using EMD, identifying the particular modes
where the information associated to desaturations appears more clearly, and
using a set of properly chosen thresholds and simple rules to count each
desaturation.

\section{Materials}
\label{sec:analisis_materials}

\subsection{Oximetry Signal}
\subsubsection{SpO$_2$ Signal Basis}
\begin{figure}
  \centering{\includegraphics[width=\columnwidth]{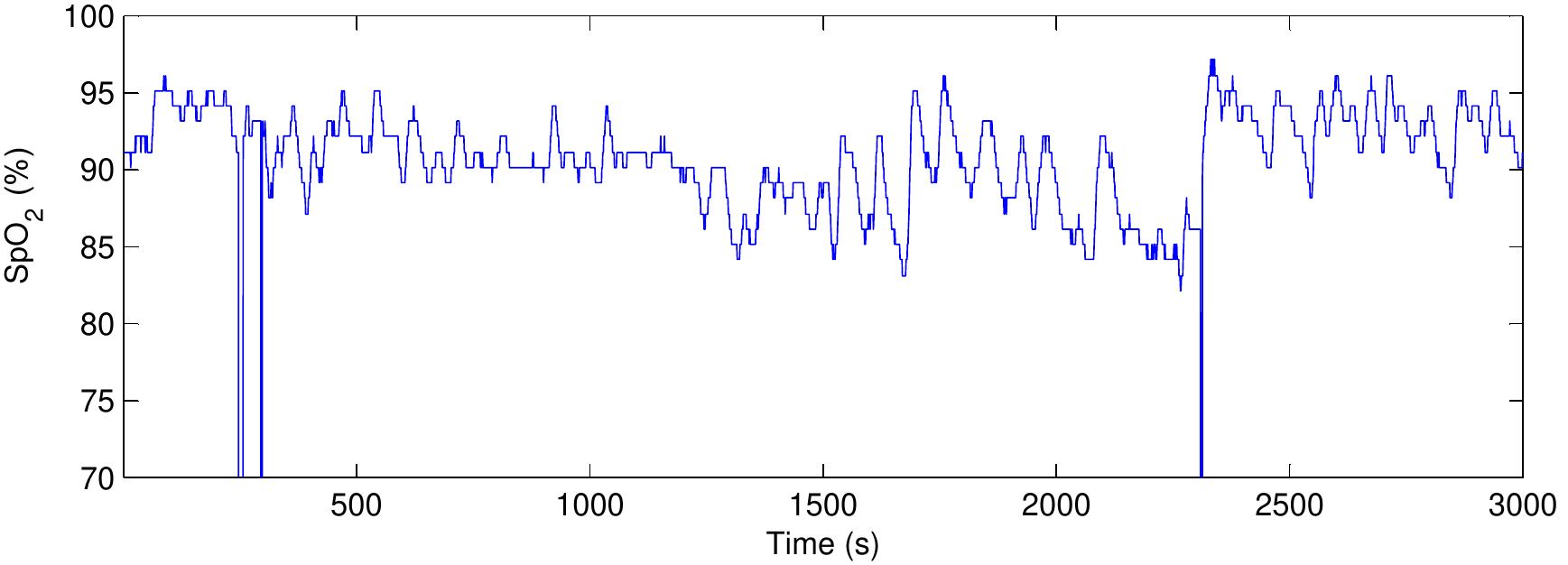}\\}
  \caption{Typical SpO$_2$ signal from a patient suffering OSAHS.}
  \label{oximetria}
\end{figure}
Oximetry is the measurement of the percent saturation of oxygen in hemoglobin.
The arterial oxygen saturation is commonly referred as SaO$_2$.
Pulse oximetry is a noninvasive estimation of the peripheral oxygen saturation
(SpO$_2$) based on the transmission, absorption, and dispersion of light as it
passes through hemoglobin. The reading is obtained using a light sensor
containing two sources of light (red and infrared) that are absorbed by
hemoglobin and transmitted through tissues to a photodetector. Measurement of
SpO$_2$ is less accurate at low values, and $70 \%$ saturation is generally
taken as the lowest accurate reading. Typical technical specifications of pulse
oximeters include a sampling rate of $1$ Hz, a resolution of $1\%$, and an
accuracy of $\pm 2\%$ in the range of $70\%$ to $100\%$.

In Fig. \ref{oximetria} a SpO$_2$ signal corresponding to a patient suffering
OSAHS is shown. The range was limited to $70 - 100\%$. Several characteristics
of this signal are illustrated in this example. Typical disconnection errors are
at $250$, $300$, and $2300$~s (the value provided by the oximeter in these
events is $0.1\%$). Examples of desaturation events can be observed at $1000$ s
and between $2000$ and $3500$ s, where sawtooth-like waveforms are present.
Additionally, a low frequency tendency can be noticed in the segment shown.

\subsubsection{SpO$_2$ and OSAHS}
A full PSG  is required for the diagnosis of OSAHS. With these records, a
specialized physician can accurately diagnose this syndrome, taking into account
the number of complete and partial obstructions (apnea and hypopnea
respectively) of breathing per hour of sleep. This quantity is known as the
Apnea-Hypopnea Index (AHI) \cite{schlosshan_sleep_2004}. It is a very expensive
study and the sleep laboratories are scarce, specially in developing countries.

The nocturnal transcutaneous pulse oximetry is used with increasing frequency
for early screenings of OSAHS due to its low cost and simplicity. During obstructive apneas, oxygen desaturations
are common, but they can be absent with hypopneas or in events with increased
upper airway resistance \cite{schlosshan_sleep_2004}. In the first case, the
desaturations show a typical sawtooth waveform with a rapid increase in SpO$_2$
during or after the arousal. However, this increase is not as abrupt in
hypopneas and the sawtooth pattern can be completely missing in central apneas.

An obstructive apnea/hypopnea event is characterized by a transient reduction or
complete cessation of breathing. In the clinical practice apneas are not
considered differently from obstructive hypopneas because these events have
similar pathophysiology. To be considered as an apnea/hypopnea event, criterion 1
or 2, plus criterion 3 of the following must be fulfilled
\cite{mcnicholas_diagnosis_2008}:

\begin{enumerate}
\item The amplitude of a valid signal related to the breathing must present a
clear decrease ($\geq50\%$) from its baseline. This baseline is defined as the
mean amplitude of the signal in stable breathing and oxygenation in the 2
minutes preceding the onset of the event.
\item A clear reduction in the amplitude of a validated measure of breathing
during sleep that does not reach the previous criterion, but occurs with an
oxygen desaturation greater or equal to $3\%$ or an arousal.
\item The duration of the event is  $10$ s or longer.
\end{enumerate}

In this work, we are focused in detecting the blood oxygen desaturations, with
the intention of identify events associated with criteria 2 and 3. Our interest
lies in estimating an index with high sensitivity for OSAHS detection. However,
as could be seen in Fig. \ref{oximetria}, this is not an easy task for real
SpO$_2$ signals. There are many problems to be solved, as artifacts,
quantization noise, baseline, etc.

\subsection{The Sleep Heart Health Study Polysomnography Database}
The Sleep Heart Health Study (SHHS) was designed to investigate the relationship
between sleep disordered breathing and cardiovascular disease \footnote{The
findings in this report were based on publicly available data
made available through the Sleep Heart Health Study (SHHS). However, the
analyses and interpretation were not reviewed by members of the SHHS and does
not reflect their approval for the accuracy of its contents or appropriateness
of analyses or interpretation.}. Polysomnograms were obtained in an unattended
setting, usually in the homes of the participants, by trained and certified
technicians \cite{redline1998methods}. The recording montage consisted of:
\begin{itemize}
   \item C3/A2 and C4/A1 EEGs, sampled at 125 Hz.
    \item Right and left EOGs, sampled at 50 Hz.
   \item Bipolar submental EMGs, sampled at 125 Hz.
    \item Thoracic and abdominal excursions  sampled at 10 Hz.
    \item Airflow (nasal-oral thermocouple), sampled at 10 Hz.
    \item Pulse oximetry, sampled at 1 Hz.
   \item  ECG  sampled at 125 Hz or 250 Hz.
   \item  Heart rate sampled at 1 Hz.
    \item Body position.
    \item Ambient light.
    \end{itemize}

Full details can be found in \cite{Quan97, Lind03}. From the conditional-use
SHHS dataset containing 1000 records, 996 were used in this work. Due to
technical reasons, four signals were discarded.

\section{Methods}
\label{sec:analisis_methods}
\subsection{Preprocessing}
The fingertip pulse oximetry signal available in the SHHS database is
complemented with information regarding the state of the oximeter. When the
patient changes its position or simply moves its limbs, this movement can
produce artifacts and render an invalid measurement, as can be seen in Fig.
\ref{oximetria}. This causes a discontinuity in the signal, with an abrupt jump
toward a saturation value of $0.1\%$. Thus, the obtained signal can have one or
more invalid portions during a study. These non-informative
components badly affect the EMD algorithm \cite{kim_extending_2012}, and they
should be avoided.

For this purpose we use the data regarding the sensor status, and we simply
eliminate the time span during which the sensor signal is invalid, with a
concatenation of the previous and posterior data. Although this may sound
unnatural, we have tried other alternatives, like interpolation using different
methods, and in all cases the interpolation also produces a perturbation in the
EMD algorithm that renders unusable the results. For this reason, we applied
this simple method.

An additional problem is related with the quantization: each quantization level
corresponds to $1\%$ of the saturation value. This quantization noise produces
artifacts in the resulting EMD decomposition. To reduce its influence and taking
into account that the desaturations produced by the apneas would have periods
larger than $5$ s, corresponding to oscillations of $0.2$ Hz, we apply a lowpass
FIR filter with a cutoff frequency of $0.25$ Hz to preprocess the signal.

\begin{figure*}[tb!]

\centering
{
\hspace{-20pt}
\subfloat[][]{
\label{fig:sub1}\includegraphics[width=0.75\columnwidth]{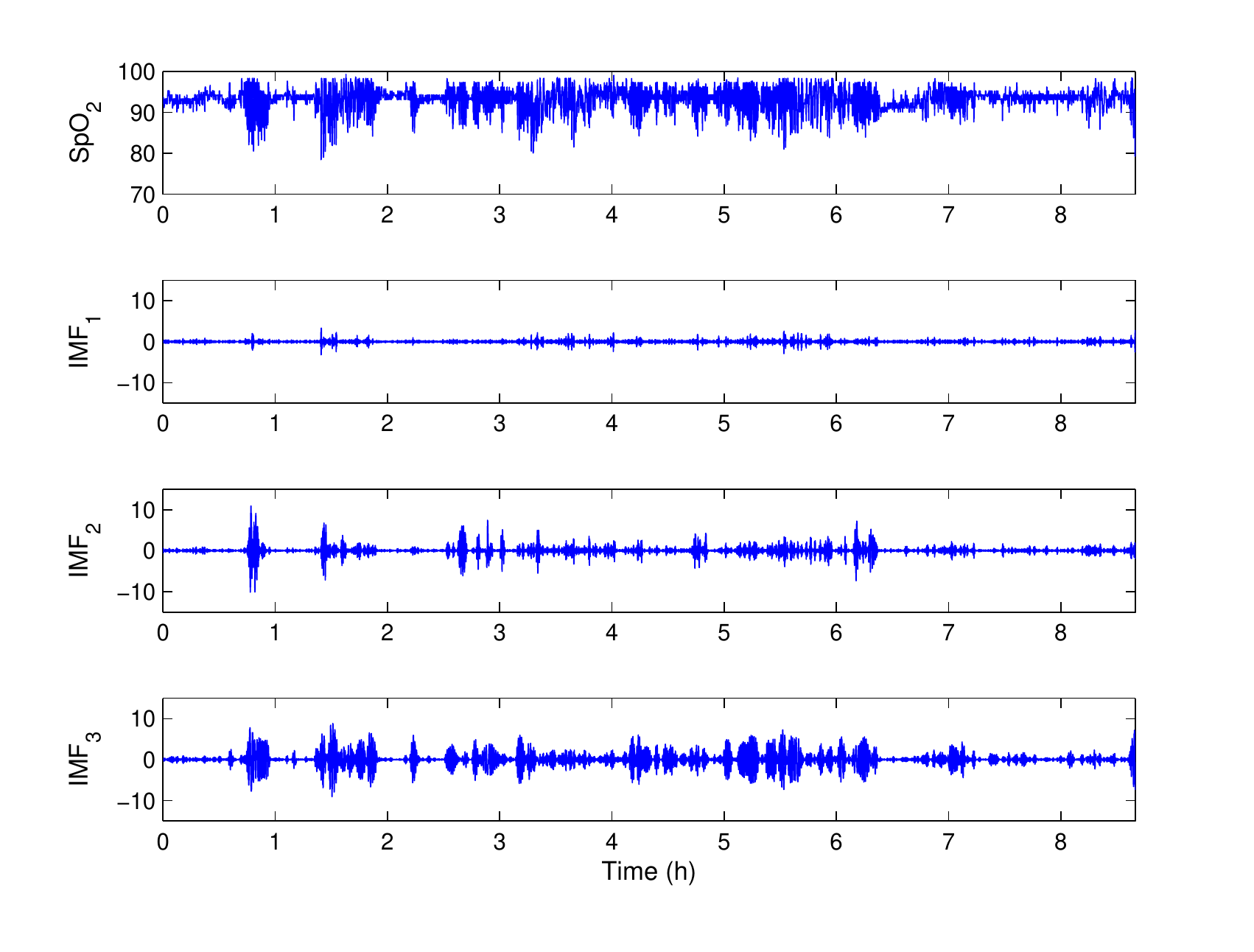}
}
\hspace{8pt}
\subfloat[][]{\label{fig:sub2}\includegraphics[width=0.75\columnwidth]{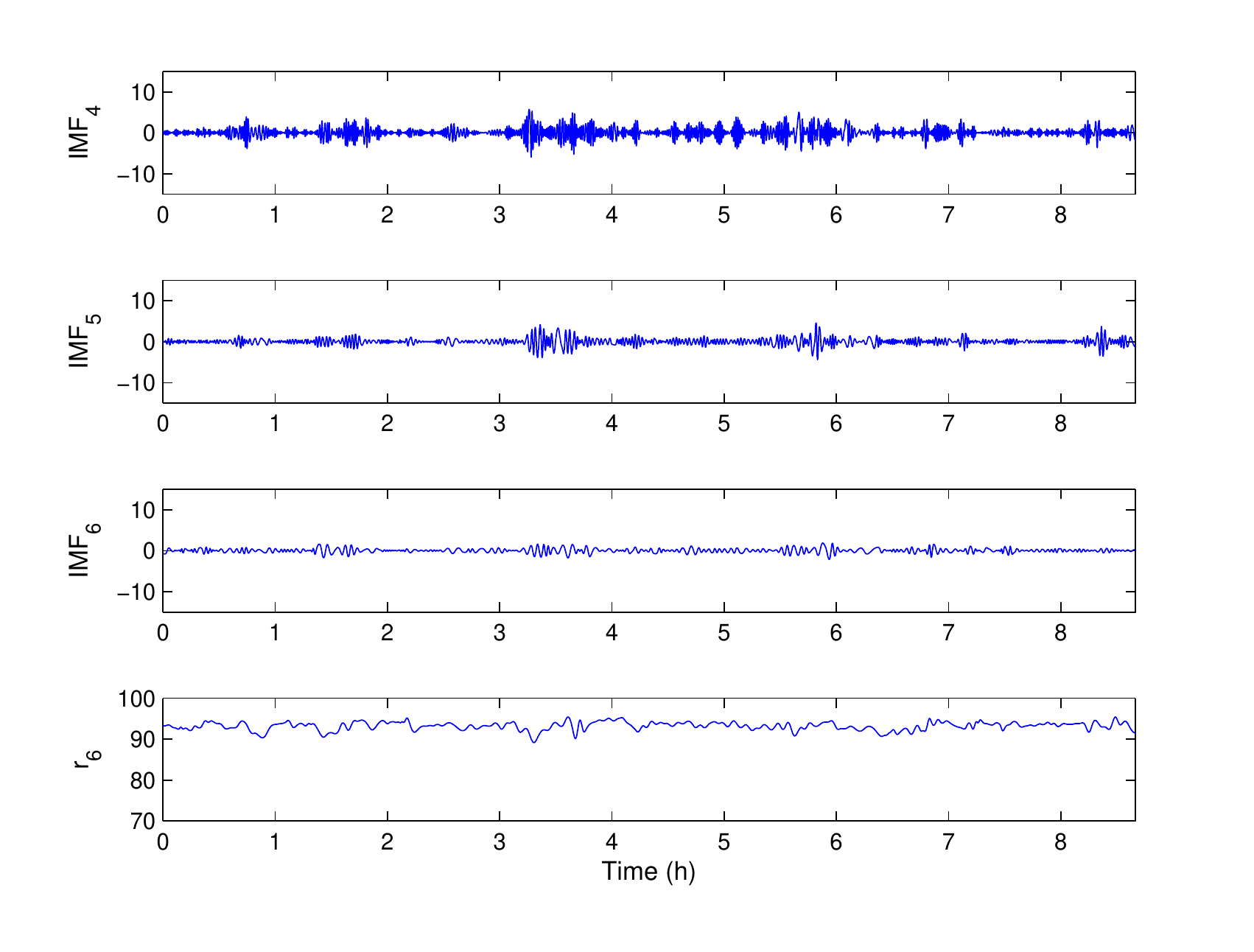}
} \\
\caption{Oxygen saturation signal and its empirical mode decomposition. (a)
Original SpO$_2$ signal (top) and IMFs 1 to 3. (b) IMFs 4 to 6 and residue.}
\label{descomposicion}
}
\end{figure*}
\subsection{EMD of SpO$_2$}
In Fig. \ref{descomposicion} the decomposition in six modes \emph{via} EMD of an
oximetry signal is shown. The signal has been preprocessed as described above.
The oximetry signal is in the upper panel in Fig.~\ref{descomposicion}a. It can
be noticed the distinctive sawtooth-like behavior of the SpO$_2$ in presence of
desaturations. The first mode of this decomposition contains the residual
quantization noise with useless information. Modes 2 to 5 seem to provide more
useful data showing oscillations where the desaturation events occur. Mode 6
and the final residue contain irrelevant information, including low-energy and
low-frequency oscillations, and the signal trend. The oscillations associated
to desaturations are distributed in different modes, making impossible to select
a single mode to detect these events. As a solution, combinations of two or more
modes are here proposed.

\subsection{Detection}
\label{sec:metodo}

We propose a method based in the EMD of the oximetry signal with the goal of
estimating an index that can be used in the screening of OSAHS.

The standard algorithms for automatic detection of desaturations are based on
the clinical criteria 2 and 3. The SpO$_2$ must  decrease at least $3\%$, and
last $10$~s or longer, to be considered as an apnea/hypopnea event. This
reduction is measured from a baseline corresponding to the normal oxygenation. Unfortunately, there is not a consensus about the
methodology for estimating this baseline \cite{berry2012rules}. One approach is
based on using the mean value of SpO$_2$ over all the study. A different method
only considers the first 3 minutes \cite{Zam1999}. Nevertheless, in some cases,
SpO$_2$ can drop to a stable value under the baseline during the sleep.

To avoid these problems, dynamic estimations of the baseline are employed.
Chiner et al. \cite{Chi1999} use the mean value of SpO$_2$ during the previous
$n$ minutes as baseline estimator. The same strategy is used by de Chazal et al.
 \cite{chazal_multimodal_2009}. Another method is applied by V\'azquez et al.
\cite{Vaz2000, burgos_real-time_2010}. In these cases, the baseline is estimated
using the top fifth percentile of SpO$_2$ values over the five minutes preceding
the event. This method do not take into account the SpO$_2$ values during
desaturations, and for this reason the baseline estimation is much more similar
to the basal value during the normal breathing. An equivalent procedure is
employed in a recent study \cite{xie_real-time_2012}, where they
adopt as baseline the mean of the top $20\%$ of the SpO$_2$ data within $1$ min.
The main failures of the algorithms for desaturation detection are related with
incorrect baseline estimations.

The new algorithm here proposed do not need these estimations. We applied EMD
to the preprocessed SpO$_2$ signal, with a maximum number of modes set to six
and a maximum number of sifting iterations set to 50. The stopping criterion was
the one proposed by Rilling \cite{Rilling03}. Auxiliary signals were obtained
by adding two, three, or four consecutive modes, considering only modes from $2$
to $6$. Each auxiliary signal was processed searching for extrema. Next, the
difference in amplitude between each local maximum and the following local
minimum ($\Delta A$), and the corresponding time interval ($\Delta T$) were
measured. If both  $\Delta A$ and $\Delta T$ are higher than certain previously
set thresholds ($\tau_{A}$ and $\tau_{T}$), a desaturation event is
detected. Finally, an oxygen desaturation index (ODI) defined as the ratio
between the number of desaturation events and the duration of the valid signal
(in hours) is calculated.

\section{Results}
\label{sec:resultados}
In the previously described algorithm, three aspects need to be experimentally
determined: the combination of EMD modes, and the parameters $\tau_{A}$ and
$\tau_{T}$. To determine these values a partition of 40 training signals was
generated. These signals were randomly selected, making a balanced training set
with 10 signals with AHI~$\leq 5$, 10 with $5< \mbox{AHI} \leq10$, 10 with $10 <
\mbox{AHI}\leq 15$ and 10 with AHI higher than $15$. This was necessary due to
the unbalance in favor of high AHI signals in the database.
The remaining signals were kept as a test database.
As an objective measure to evaluate the algorithms, we use the area under the
ROC curve (AUC) \cite{Fawcett2006}. This measure allows for a comparison of
different classifiers for the whole range of threshold. A bootstrap estimator
\cite{Efron1986} of the AUC using 100 replicates was applied and the confidence
intervals were estimated \cite{Macskassy2004}.

\subsection{Parameter selection}
To find the best combination of modes and thresholds $\tau_{A}$ and $\tau_{T}$,
a series of experiments was performed over a training dataset with the set of 40
randomly selected signals. The values of $\tau_{A}$ were varied from
$\tau_{A}=1$ to $\tau_{A}=4$ in steps of $0.1$, $\tau_{T}$ was varied from
$\tau_{T}=10$ to $\tau_{T}=30$ in steps of $1$, and the ODI was
estimated for each signal. The signals were separated into two classes using as
thresholds the polisomnography based AHIs $= 5$, $10$ and $15$.

\begin{figure}
  \centering{\includegraphics[width=\columnwidth]{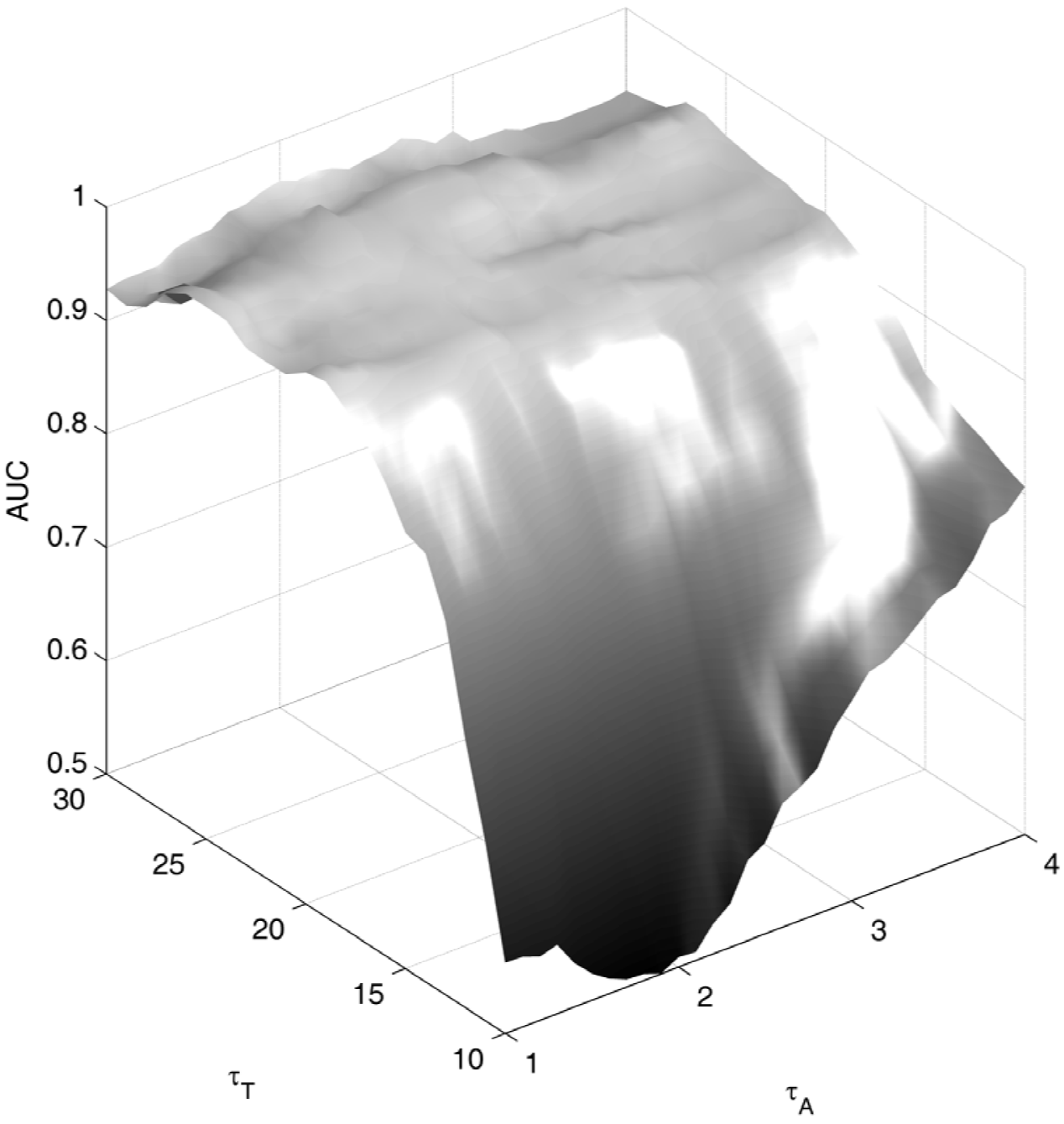}\\}
  \caption{Parameter selection using AUC. Diagnostic threshold for AHI=15.}
  \label{fig3d}
\end{figure}

All mode combinations were explored. The behaviors were qualitatively similar.
The combination of modes that yielded the best results was the sum of modes 3,
4, and 5. In Fig. \ref{fig3d} we show the AUC for a threshold AHI $=15$ as a
function of $\tau_{A}$ and $\tau_{T}$. The main reason for this combination
of modes been more effective that using a single mode is that the oscillations
more related with the desaturations events are present in one of these three modes.
These events can not be captured in a single mode because its amplitude and
duration are changing in time and among patients.

In Fig. \ref{fig3d} it can be noticed that the best AUC values are in two
well-localized ``ridges'': one corresponding to $\tau_T$ around $19$--$20$ and
the other in $\tau_T$ with values rounding $24$--$25$. The maximum is AUC
$=0.972$ for $\tau_T=19$ and $\tau_A=1.1$.

\subsection{Evaluation on the test database}

The proposed algorithm was applied to the $70\%$ of the
remaining signals in the database (669 cases) using the best combination of
modes and the optimal values of the parameters.
The method was compared with the two mentioned baselines \cite{Chi1999,
Vaz2000}.
Three values for the reference diagnostic AHI threshold were used. Table
\ref{tab:AUC5} displays the results for a threshold of AHI$=5$, $10$, and $15$.
The estimated value of AUC and the $90\%$ confidence intervals are shown.

\begin{table}[!t]
    \caption{AUC for the different detection algorithms. AUC$_{min}$ and
AUC$_{max}$ indicate the limits of the $90\%$ confidence intervals.}
    \label{tab:AUC5}
    \centering
   \begin{tabular}{l|c|c|c|c}
    \hline
    Method                                             & AHI$_{thr}$          &
{AUC}  & AUC$_{min}$     & AUC$_{max}$ \\
    \hline
    \hline
                                                       & 5                    &
0.687  & 0.526            & 0.872 \\
   Chiner et al. \cite{Chi1999}                        & 10                   &
0.754  & 0.694            & 0.798 \\
                                                       & 15                   &
0.749  & 0.701            & 0.795 \\ \hline
                                                       & 5                    &
0.856  & 0.723            & 0.941 \\
   V\'azquez et al. \cite{Vaz2000}                     & 10                   &
0.894  & 0.854            & 0.922 \\
                                                       & 15                   &
0.905  & 0.880            & 0.920 \\ \hline
                                                       & 5                    &
0.888  & 0.837            & 0.962 \\
    EMD                                                & 10                   &
0.912  & 0.879            & 0.941 \\
                                                       & 15                   &
0.923  & 0.898            & 0.942 \\
    \hline
   \end{tabular}
\end{table}

\begin{figure}
  \centering{\includegraphics[width=\columnwidth]{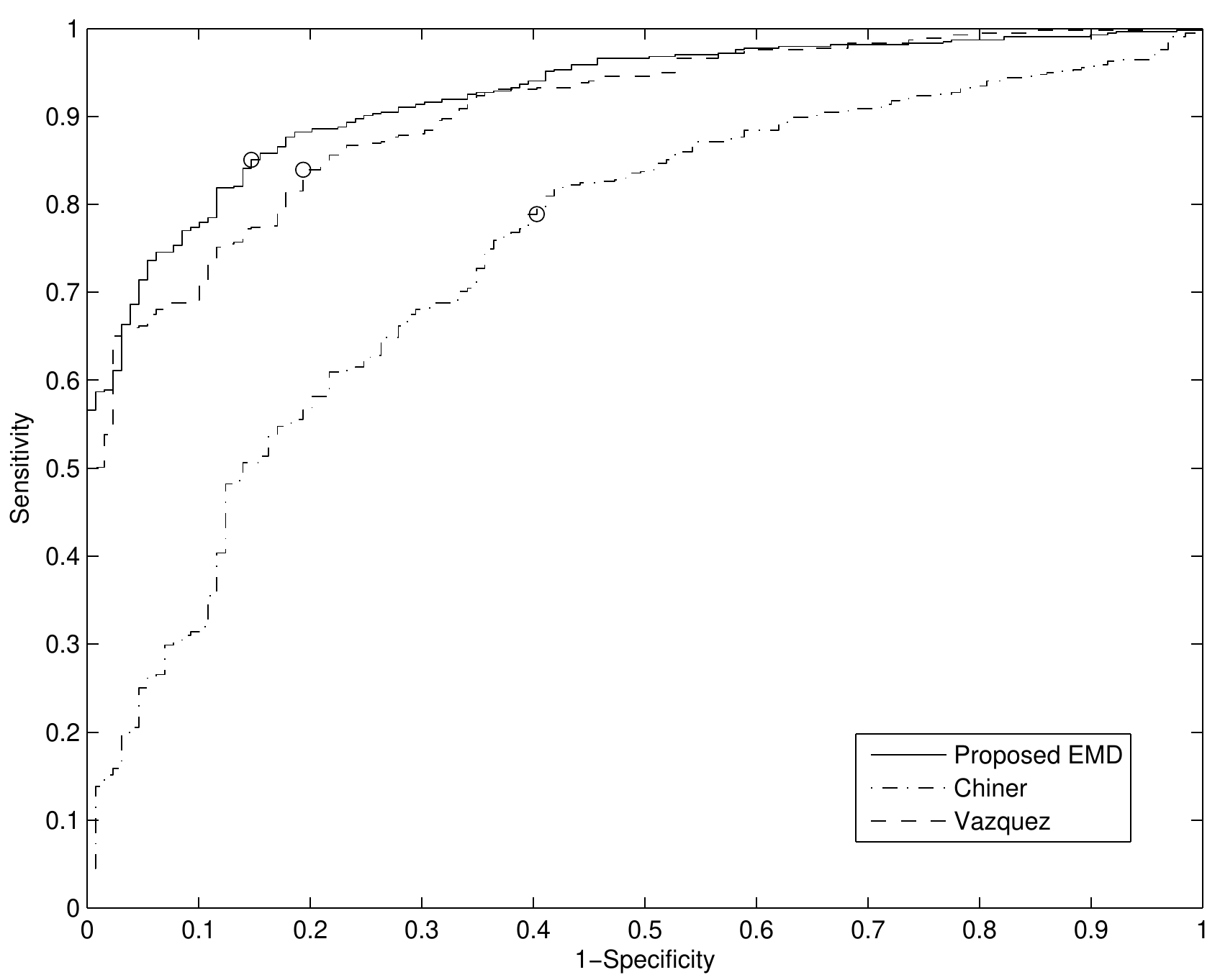}\\}
  \caption{ROC curve for the proposed method. Diagnostic threshold for
reference
AHI $=15$. For each curve, the optimal operating point is marked with a circle.}
  \label{fig:ROC}
\end{figure}

It can be seen that the AUC for the proposed method is the best among the
tested alternatives.
In case of the reference AHI threshold of 15, the resulting ROC curve can be
seen in Fig. \ref{fig:ROC}. The circle shows the optimal operating point which
maximizes both the sensitivity ($se$) and the specificity ($sp$).
This point corresponds to a diagnostic threshold of ODI $\tau_D=18.512$, and produces
$se=0.851$ and $sp=0.853$. The figure also shows the ROC curve and
optimal operating point for the algorithm by Chiner \cite{Chi1999} with
$se=0.789$ and $sp=0.597$ for $\tau_D=3.095$, and the
method by V\'azquez \cite{Vaz2000} with $se=0.839$
and $sp=0.806$ using a $\tau_D=11.351$.

The final test was done using the remaining $30\%$ of the
database (287 signals never used in previous stages) estimating
$se$ and $sp$ at the optimal operating points of the three analyzed methods. The sensitivity and specificity
corresponding
to the here proposed method were $0.838$ and $0.855$ respectively. In the case
of the algorithm by Chiner et al., $se$ and $sp$
were $0.812$ and $0.618$ respectively, and the corresponding ones for the method by Vazquez
were $0.829$ and $0.818$.

\section{Discussion}
\label{sec:discussion}
The utilization of pulse oximetry as a sole signal to assist OSAHS diagnosis is
still a controversial issue. Collop et al. \cite{collop2007clinical} stated that 1 or 2 channels (including oximetry)
home-unattended studies had wide variance of false positives, and that the
evidence to support these studies to make a diagnosis of OSAHS is insufficient.
Analysis of nocturnal oximetry has been applied as a potential diagnostic
screening tool over the two past decades, but the signal interpretation were
highly dependent on the physician, and on the technical performance. Recently it
was demonstrated that, when treated with appropriated and sophisticated
algorithms, overnight oximetry recording appears to be a very sensitive and
specific screening method of OSAHS \cite{poupard2012novel}. Pulse oximetry is
accepted as the sole diagnostic evaluation criterion in United States, Australia
and Sweden \cite{boehning}. The Apnea Task Group of the German Society for Sleep
Research and Sleep Medicine (DGSM) has stated that pulse oximetry can be
employed to attain a tentative diagnosis that requires further evaluation at a
sleep laboratory.

The results of pulse oximetry can be limited by artifacts due
to inaccurate readings (especially in obese patients), hypotension, and
abnormalities in the hemoglobin, among several factors. These drawbacks make
evident the need of signal processing and pattern recognition techniques in
order to detect and reduce the effects of noise and artifacts. In previous
methods based on oxygen saturation obtained by pulse oximetry, sensitivity and
specificity ranged from $31$ to $98\%$, and from $41$ to $100\%$ respectively,
according to \cite{Hor2007, Ray2003, Vaz2000}. This high variability is caused
by the differences among the devices, populations, and the applied signal
processing methods \cite{Net2001}. The results of the here presented method overcome the ones of \cite{Vaz2000,Chi1999}, as shown in the previous section.
Additionally, the database here used (996 patients) is larger
than those utilized in \cite{Vaz2000} (241 patients) and \cite{Chi1999} (275
patients), which may explain the discrepancies among the results of the original references and the obtained in our work.
 Given that our results were obtained using all the methods over the same larger database, the proposal of this work clearly outperforms the analyzed alternatives. Additionally, the bootstrapping approach allowed us to estimate the confidence intervals of the AUC, which was not done in the other cases. This is a rigorous methodology which, to our knowledge, was not used in this area in previous works.

As the ROC curves for the proposed approach are above the ones corresponding to
the standard methods in the whole range, this new technique produces a
better compromise between sensitivity and specificity.

One limitation of our method, as in all methods
based only on desaturation, is that there is no information regarding the sleep
stage of the patient. The number of desaturations associated to apnea by hour of sleep is impossible to
estimate without knowing if the patient is asleep or not. Another limitation may be related to the signal quality. As above-mentioned, if the signal has artifacts related to movements or disconnections, that
segments are eliminated prior to the EMD. Thus, for this method
to be valuable, the signal quality must be assessed and low quality studies
should be discarded.

\section{Conclusion}
\label{sec:conclusiones}
A new algorithm for SpO$_2$ signal analysis using EMD was proposed. It was shown
that the information from desaturations was mainly concentrated in EMD modes
2-5. Based on this information, a detection algorithm using a combination of
these modes was proposed. The optimal parameters were determined using a
balanced training database. This desaturation detector was used to produce an
ODI that is here used to detect OSAHS. It was found that the best alternative
was to combine modes 3, 4 and 5. When compared AUC over the test database with
the two standard algorithms, it was seen that the here proposed method
outperforms the standard ones, with narrower confidence intervals.
As future work, we are interested in testing more advanced methods for EMD that
avoid the problem of mode mixing, to improve even more these results.
Furthermore, a more balanced database would enable a better parameter selection.

\section*{Acknowledgments}
Competing interest: None declared.

Funding: This work was supported by the National Agency for Scientific and
Technological Promotion (ANPCyT) under Grants PAE--PID--2007--00113,
PICT-2012-2954, Universidad Nacional del Litoral under projects UNL-CAI+D R4-N14
and UNL CAI+D 2011 58-519, Universidad Nacional de Entre R\'{\i}os, and the
National Council on Scientific and Technical Research (CONICET).

Ethical approval: Not required.

The authors thank Carlos Pais from CardioCom S.R.L. for his support during the
realization of this work.


\end{document}